\documentclass{aa}
\usepackage{amssymb,epsfig,graphicx}
\usepackage{natbib}
\usepackage{longtable}
\usepackage{txfonts}
\begin{document}
   \title{The host of GRB 060206: kinematics of a distant galaxy}

\author{Christina C. Th\"one \inst{1} \and Klaas Wiersema \inst{2} \and C\'edric Ledoux \inst{3} \and Rhaana L. C. Starling \inst{4} \and Antonio de Ugarte Postigo \inst{3,5} Andrew J. Levan \inst{6} \and Johan P. U. Fynbo \inst{1}  \and Peter A. Curran \inst{2}   \and Javier Gorosabel \inst{5} \and Alexander J. van der Horst \inst{2} \and Alvaro LLorente \inst{7} \and Evert Rol \inst{4}  \and Nial R. Tanvir \inst{4}  \and Paul M. Vreeswijk \inst{1,3} \and Ralph A. M. J. Wijers \inst{2} \and Lisa J. Kewley \inst{8}}          


\institute{Dark Cosmology Centre, Niels Bohr Institute, University of
	  Copenhagen, \mbox{Juliane Maries Vej 30, 2100 Copenhagen, Denmark}
	 \and
	  Astronomical Institute `Anton Pannekoek', University of Amsterdam,
	  Kruislaan 403, 1098 SJ Amsterdam, the Netherlands
	  \and
European Southern Observatory, Alonso de C\'ordova 3107, Vitacura, Santiago 19, Chile
\and
Dept. of Physics and Astronomy, University of Leicester, University Road,
Leicester LE1 7RH, UK
\and
Instituto de Astrof\' isica de Andaluc\' ia, (IAA-CSIC), Apdo. 3004, E-18080 Granada, Spain
\and
Department of Physics, University of Warwick, Coventry CV4 7AL, UK
\and
XMM-Newton Science Operations Centre, European Space Agency, Villafranca del Castillo, PO Box 50727, 28080 Madrid, Spain
\and
University of Hawaii, 2680 Woodlawn Drive, Honolulu, HI 96822, USA}	
  
\offprints{cthoene@dark-cosmology.dk}

   \date{Received 2007; accepted 2008}
 
  \abstract
   {GRB afterglow spectra are sensitive probes of interstellar matter along the line-of-sight in their host galaxies as well as intervening galaxies. The rapid fading of GRBs makes it very difficult to obtain spectra of sufficient resolution and S/N to allow for these kinds of studies.}
   {We investigate the state and properties of the interstellar medium in the host of GRB\,060206 at $z= 4.048$ with a detailed study of groundstate and fine-structure absorption lines in an early afterglow spectrum. This allows us to derive conclusions on the nature and origin of the absorbing structures and their connection to the host galaxy and/or the GRB. 
   }
   {We use early (starting 1.6h after the burst) WHT/ISIS optical spectroscopy of the afterglow of the gamma-ray burst GRB\,060206 detecting a range of metal absorption lines and their fine-structure transitions. Additional information is provided by the afterglow lightcurve. The resolution and wavelength range of the spectra and the bright afterglow facilitate a detailed study and fitting of the absorption line systems in order to derive column densities. We also use deep imaging to detect the host galaxy and probe the nature of an intervening system at $z = 1.48$ seen in absorption in the afterglow spectra.}
   {We detect four discrete velocity systems in the resonant metal absorption lines, best explained by shells within and/or around the host created by starburst winds. The fine-structure lines have no less than three components with strengths decreasing from the redmost components. We therefore suggest that the fine-structure lines are best explained to be produced by UV pumping from which follows that the redmost component is the one closest to the burst where also \ion{N}{V} was detected. 
   The host is detected in deep {\emph HST} imaging with F814W$_{AB}$ = 27.48 $\pm$ 0.19 mag and a 3 $\sigma$ upper limit of $H = 20.6$ mag (Vega) is achieved. A candidate counterpart for the intervening absorption system is detected as well, which is quite exceptional for an absorber in the sightline towards a GRB afterglow. The intervening system shows no temporal evolution as claimed by Hao et al. (2007) which we prove from our WHT spectra taken before and Subaru spectra taken during the observations from Hao et al.}
  {}


   \maketitle
%

\section{Introduction}

In recent years, GRBs have been detected out to very high redshifts with currently four at $z>5$ \citep{Jakobsson06a,cenko06,ruiz-velasco, Haislip05, Kawai06}. Theory and observations suggest that $\sim$10\% of {\it Swift}-detected gamma-ray bursts (GRBs) will originate at redshifts $z \gtrsim 5$ and $\sim$3\% at redshifts $z \gtrsim 6$ \citep{Bromm06, Daigne06, Tanvir07}. Furthermore, the observed flux is not expected to fade significantly with increasing redshift as time dilation effects lead to earlier observations of the afterglow in the GRB  restframe. Hence GRBs should be detectable out to $z \gtrsim 10$ \citep{Ciardi00} making them the most distant observable objects in the Universe. Indeed, some of the most distant GRB afterglows were also among the most luminous ones observed e.g. GRB\,050904, $z = 6.3$ \citep{Kawai06} and GRB\,060206, $z = 4.048$, presented here \citep[see also][]{Monfardini06}. As such, these events can teach us something about galaxy evolution and the early Universe: the occurrence of GRBs in starforming regions in high redshift galaxies allows us to probe sightlines to these galaxies with high spectral resolution, by exploiting the brightness and simple spectral slope of GRB afterglows \citep[e.g.][]{ProchaskaISM, Delia, Vreeswijk07}. 

Through studies of absorption lines in the spectra of GRB afterglows, it is possible to determine properties of these distant galaxies which are otherwise hard to obtain. One key piece of information that can be derived from resonant metal absorption lines is the metallicity along the line-of-sight towards the GRB in the host galaxies. These restframe-UV transitions are shifted into the optical regime above $z\sim2$ and hence observable with ground-based telescopes. The line-of-sight metallicities derived from afterglow spectroscopy are higher than the mean QSO absorber metallicity as a function of redshift \citep{fynbob,ProchaskaISM}. This might, however, only be caused by a selection effect as the sightlines probed by QSOs and GRBs are probably different \citep[][see Fig. 1]{Fynbo08, ProchaskaISM}. At redshifts below $z\sim1$ GRB hosts have lower metallicities than a sample of blue, star-forming galaxies from the SDSS survey \citep{Tremonti}. For these nearby galaxies, the metallicities are derived from emission line diagnostics and reflect the metallicity of the HII regions within these galaxies \citep{Savaglioreview}.

QSO absorption systems have been used for a long time to study distant galaxies in a similar way as done with GRBs, but an association of the absorbing system with a galaxy is not always possible due to the brightness of the QSO itself. Furthermore, QSO sightlines seem to probe the less dense outer regions of galaxies, whereas GRBs often reside inside a galaxy and their sightlines also allow us to probe the inner parts of the galaxy and the immediate environment of the burst \citep[see e.g. Fig. 1 in ][]{ProchaskaISM}. In all GRB spectra with the right wavelength coverage and high enough S/N and resolution, fine-structure lines such as \ion{Si}{II*}, \ion{Fe}{II*}, \ion{O}{I*} \citep{ProchaskaISM} and even higher levels have been detected \citep[e.g.][]{Vreeswijk07}. These lines are produced by collisional excitation or indirect UV pumping of the levels from the strong UV radiation field of the GRB and may therefore be closely connected to the GRB itself. In QSO absorbers, only fine-structure lines of \ion{C}{I} and \ion{C}{II} have been detected so far \citep[][and references therein]{Silva}, but no other fine-structure transitions such as \ion{Si}{II*} or \ion{Fe}{II*}.

In our first paper on the dataset presented here \citep[][hereafter Paper I]{fynbob} we showed initial results of the WHT and NOT spectra of the afterglow of  GRB\,060206, focussing on the metallicity which we compared to a large sample of QSO absorber metallicities and the sample of known GRB afterglow metallicities. In this second paper we analyze the full set of absorption lines, and add a more detailed analysis of the properties of the host from the photometric detection of a possible host galaxy and broadband afterglow properties. The paper is organized as follows: In \S2 we present the spectroscopy of the GRB afterglow and the observations of the host galaxy. In \S3 we describe the  absorption line measurement method and results. In \S4 we analyze the discrete velocity components in the host through the detected (fine structure) lines and the line-of-sight extinction. We present the properties of the host galaxy in emission in \S5 through late deep imaging. In \S6 we discuss the absorption line properties of the intervening 
absorbers, and evaluate the possibility that a galaxy close to the afterglow location is responsible for one of the intervening absorber systems. Throughout this paper we use the cosmological parameters H$_{0}$ = 70 km s$^{-1}$Mpc$^{-1}$, $\Omega_{\rm M}$ =
0.3 and $\Omega_\Lambda$ = 0.7.


\section{Observations}
\subsection{Spectroscopy}
We make extensive use in this paper of the dataset described in Paper I, to which we refer for details on the observations and data reduction. 
For clarity, we reiterate here the important steps. We obtained a medium
resolution spectrum with the Intermediate-dispersion Spectroscopic and Imaging
System (ISIS) on the 4.2m William Herschel Telescope (WHT) on 2006 February 6,
starting 1.61 hours after the burst.  We took two spectra with exposure times
of 900 s each, separated only by the read-out time of the CCD. The
spectrograph has a blue and a red arm, providing FWHM resolutions of 1.68 and 0.82
\AA, respectively. A redshift of $z=4.04795\pm0.00020$ was determined from
several absorption lines within the host (Paper I). At this redshift, the
WHT spectra cover the rest frame wavelength range of approximately 753 -- 1030 and 1228 -- 1387 \AA\ for the blue and red arm, respectively. As the Ly$\alpha$ line falls into the gap between the two arms, we used the column density derived from the NOT spectra (Paper I) for deriving metallicities. The signal-to-noise ratios of the spectra are high, strongly aided by the sudden, strong rebrightening of the afterglow just prior to and during the spectroscopic
observations reaching a maximum of R$\sim$ 16.4 mag \citep{Wozniak06}.            

Furthermore, we obtained 8 $\times$ 1800s low-resolution (R $\sim$ 300) spectra from the SUBARU archive \cite{baba02} from about 6 to 10 hrs after the burst. We reduced the spectra with standard tasks in IRAF.        

The optical afterglow of this burst proved bright with R$\sim$ 18.1 mag after 1 day, and owing to its fortunate location on the sky (RA 13:31:43, DEC +35:03:03 (J2000)), it was observed at a large range of Northern observatories \citep[e.g.][]{Wozniak06,Monfardini06,Stanek07}. A detailed discussion of photometric afterglow observations and the lightcurve physics can be found in \cite{Curran07}. We use these observations together with Swift XRT data to analyze the broadband optical to X-ray extinction (see Sec.\ref{extinction}).

\subsection{Imaging}
We observed the location of the afterglow on February 16, 2007, over a year after the burst, using the GMOS-N instrument on Gemini-North with the aim of detecting the host and intervening absorber galaxies in emission. Observations were performed in the GMOS $r^\prime$ band, which has a similar shape and throughput as the SDSS $r^\prime$ band. The dataset consists of 9 $\times$ 300 s exposures with a seeing of $\sim$1 arcsec and an airmass of $\sim1.1$. Reduction was done using the \emph{Gemini} package in IRAF and the photometry was calibrated against archival SDSS observations of the GRB\,060206 field \citep{CoolGCN}. For the combined image we derive a 3$\sigma$ upper limit of $r^\prime_{AB}$=  26.3 which was obtained by examining the standard deviation of 50 blank apertures placed at random positions on the image.

We also observed the GRB field in the near-infrared on June 3, 2007 using the Omega2000 instrument on the 3.5m telescope at the Calar Alto Astronomical Observatory (CAHA). Observations consisting of 50 $\times$ 60 s were performed in the {\it H}-band with a limiting magnitude of $H$=20.6 mag (Vega).

GRB 060206 was finally observed with {\it HST} on 25 November 2006, using the ACS/WFC and F814W filter under the program 10817 (PI H.-W. Chen), which are now publicly available from the {\it HST} archive. 8 dithered exposures were obtained with a total exposure time of 9886 s. We processed these data through the standard
multidrizzle pipeline, setting the linear drop size ({\tt pixfrac}) to one, and the
final output scale to 0\farcs033 per pixel. Performing relative astrometry between our early NOT observations \citep{fynbob,Curran07} allows us to place the afterglow on this image with an accuracy of $\sim 0\farcs05$.

\section{Extinction along the line-of-sight}\label{extinction}

To obtain limits on the absorption along the line of sight to GRB\,060206, we fit the near-infrared to X-ray spectral energy distribution (SED) in count space \citep{Starling07}, and using the metallicity as determined through the sulphur lines of
[S/H] = -0.86. The SED is created using RJHK$_{\rm S}$ photometry and an X-ray spectrum from {\it Swift} XRT \citep{Gehrels,Burrows}, centred at $\sim$2.8 hours after burst, detailed in \cite{Curran07}. We note that this is close in time to our WHT spectra, which
were taken at mid-time 1.86\,hrs after burst. We perform fits using an absorbed single power law, and then a broken power law to model a possible cooling break.

The SED is well fit with a single power law with  $\beta = 0.93 \pm 0.01$ and $\chi^2$/dof = 1.05. A broken power law does not provide a significant improvement in the fit according to the $F$-test. We find no evidence for intrinsic ($z=4.048$) optical or X-ray extinction
above the Galactic values. At this high redshift of $z=4.048$ the metal edges that dominate the X-ray extinction are shifted out of the XRT energy range and are not sensitively probed. The near-infrared and optical data at this redshift, however, probe restframe
UV and blue bands, and are therefore sensitive to extinction. We estimate a 3$\sigma$ upper limit for the intrinsic optical/UV extinction of $E(B-V) < 0.01$. To derive this we fitted the SED in count space from X-ray to optical simultaneously \citep[as described in][]{Starling07} with an LMC, SMC and MW extinction law and increased the value of E(B-V) until $\chi^2 = 1$, corresponding to a 1$\sigma$ deviation. This shows that the intervening systems as well as the host velocity systems have low dust content, which can be used to distinguish between different origins of these systems.

\section{Absorption line analysis}\label{abs}

\begin{figure}
\centering
\includegraphics[width=7cm]{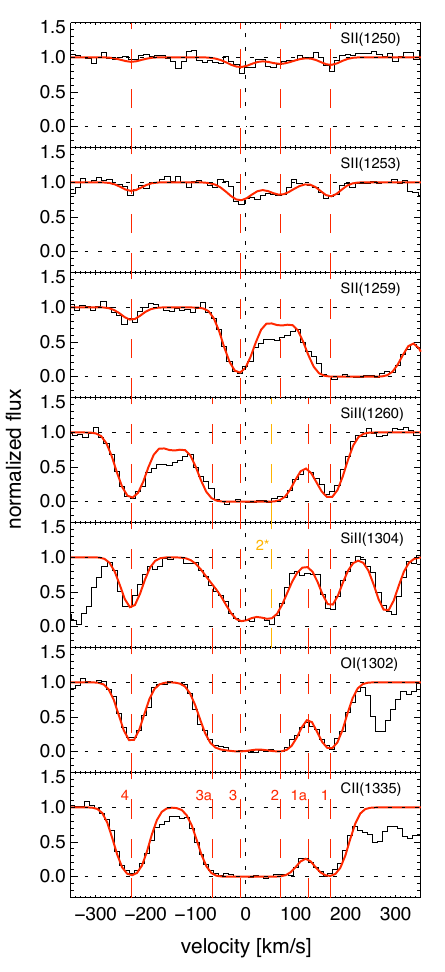}
   \caption{Fits to the four velocity components (plus two subcomponents) of the resonant absorption transitions in the host galaxy. The dashed red lines indicate the different velocity
     components as denoted in the bottom panel, the dotted line is the adopted zero velocity corresponding to $z=4.048$. Note that the second component of \ion{Si}{II} marked in orange and denoted as 2* is at a slightly different position as for the other lines. The \ion{S}{II}\,$\lambda$1259 is blended with \ion{Si}{II}\,$\lambda$1260 and was excluded from the common \ion{S}{II} multiplet fit, for the plot of the $\lambda$1259 line we therefore adopt the column derived from the fit of $\lambda\lambda$1250 and 1253. The components and their derived properties are listed in Table \ref{abslines}.}
\label{allines}
\end{figure}

\begin{figure}
\centering
\includegraphics[width=7.2cm]{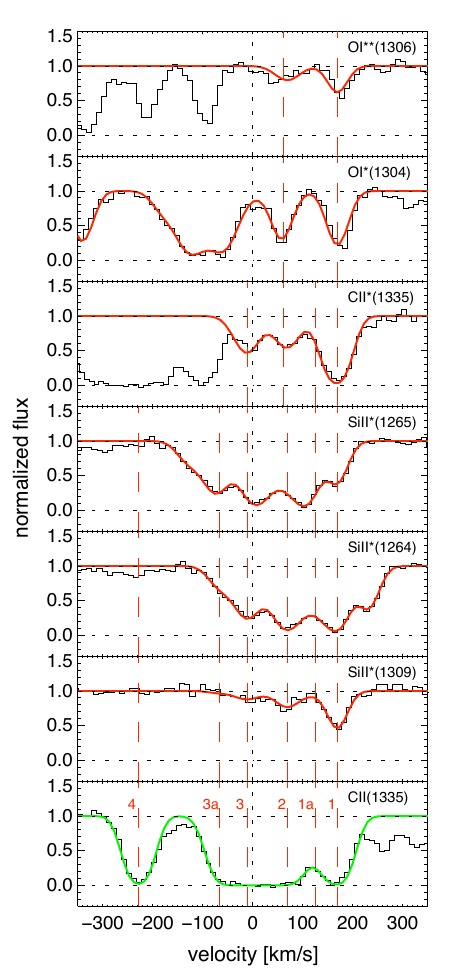}
   \caption{Same as Fig. \ref{allines} for the fine-structure lines detected, the bottom panel shows the fit to the \ion{C}{II} system for comparison. The fine-structure lines only show three or less components. \ion{O}{I*}$\lambda$1304 is blended with
     \ion{Si}{II}\,$\lambda$1304 and only the two redmost component have been
     fitted for both \ion{O}{I**}\,$\lambda$1304 and
     \ion{O}{I**}\,$\lambda$1306. \ion{Si}{II*}\,$\lambda\lambda$ 1264 and 1265
     and \ion{C}{II*}\,$\lambda\lambda$ 1335.6, 1335.7 are intrinsically blended
     and have been fitted together - for clarity, only the fit to \ion{C}{II*}$\lambda$ 1335.6 is shown.}
\label{finelines}
\end{figure}

\begin{figure}
   \centering
   \includegraphics[width=7cm]{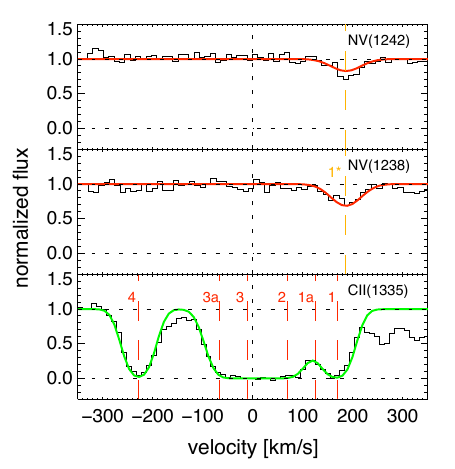}
      \caption{Fit to the \ion{N}{V} doublet at $\lambda\lambda$1238,1242 and comparison to the components in \ion{C}{II}. The only component detected, named 1*, is at a slightly higher redshift than comp. 1 in all other lines which might indicate a different origin in the host galaxy.}
         \label{NV}
   \end{figure}

\begin{figure}
\centering
\includegraphics[width=7cm]{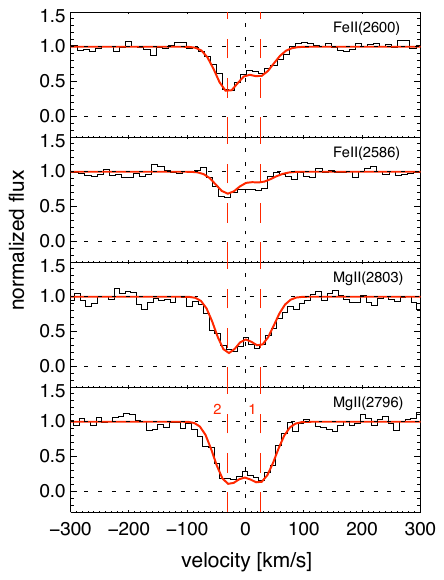}
   \caption{Fit to the intervening system in the line of sight at redshift $z=1.48$. The dotted lines indicate zero velocity which is the average of the velocity components. The two \ion{Mg}{II} and \ion{Fe}{II} transitions show two velocity components indicated by the dashed lines which have been fitted as doublets.}
      \label{interveningfit}
\end{figure}

We reported equivalent widths and column densities for a selection of absorption lines in Paper I, with the aims of deriving a metallicity for the line-of-sight within the host, the exact velocity decomposition of the systems and the fine-structure line ratios as a function of velocity.  In this paper we present a full analysis of all absorption lines in the host galaxy and the intervening systems. The results of the host galaxy systems are presented in Table~\ref{abslines} and Table~\ref{abslinesfine} as well as Fig.~\ref{allines}, \ref{finelines} and \ref{NV}, the fits for the intervening system can be found in Table~\ref{abslinesinter} and Fig.~\ref{interveningfit}.

For the host galaxy with a systemic redshift of $z=4.048$, we detect four main velocity systems at $v_1=$~169\,km\,s~$^{-1}$, $v_2=$~69\,km\,s~$^{-1}$, $v_3=$~--11\,km\,s~$^{-1}$ and $v_4=$~--228\,km\,s~$^{-1}$ which we refer to as systems 1 to 4 respectively. In order to obtain a better fit, we had to introduce two additional subcomponents for all strong lines (see below) which are at velocities of $v_{1a}=$~127\,km\,s~$^{-1}$ and $v_{3a}=$~--67\,km\,s~$^{-1}$. In addition, we find an intervening system at $z~=~1.47895$ which shows two components at $v_1=$~25\,km\,s~$^{-1}$ and $v_1=$~--30\,km\,s~$^{-1}$. Zero velocities are chosen arbitrarily as the midpoint of the entire system, as the exact redshift of the host and the intervening galaxies are unknown.

In order to fit the different components, we use the FITLYMAN package in MIDAS \citep{FontanaBallester}. The fitting is, however, complicated by the strong blending as well as saturation in a number of lines. We therefore use the weak, unsaturated \ion{S}{II}\,$\lambda$1253 transition as a template for the position of the four main components $v_1$ to $v_4$. For the Si-lines as well as \ion{O}{I} and \ion{C}{II} and the corresponding fine-structure lines, we need two additional components 1a and 3a in the blue wings of the saturated components 1 and 3 in order to achieve a good fit. In order to obtain a reasonable fit, we had to put component 2 at a slightly different redshift for \ion{Si}{II} with a velocity difference of $\Delta v_{2-2*} = $18~km~s$^{-1}$, which is therefore denoted as component 2*.

Multiplet transitions were fitted such that for each component all individual transitions have the same column density. This is useful in those cases where one of the transitions is blended with another line in order to disentangle the contributions from the different elements to one absorption feature. Blended lines were fitted together such as \ion{C}{II*}\,$\lambda$1335.6 and \ion{C}{II*}\,$\lambda$1335.7 and \ion{Si}{II*}\,$\lambda\lambda$1264 and 1265. In case of the blended lines \ion{Si}{II}\,$\lambda$1260 and \ion{S}{II}\,$\lambda$1259, \ion{S}{II} was excluded for the common fit of the \ion{S}{II} lines. For \ion{Si}{II} and \ion{O}{I*}\,$\lambda$ 1304 that were fitted together, the contribution from \ion{O}{I*} is negligible and only 2 components were fitted for the fine-structure line and the same for \ion{O}{I**}. For the fine-structure lines \ion{C}{II*}\,$\lambda\lambda$1335.6,1335.7, we fitted 3 components that are not blended with the ground state transition. The fitted values for the different components are then listed in Table \ref{abslines} for the resonant transitions and Table \ref{abslinesfine} for the fine-structure lines.

As shown in Fig. \ref{allines}, the resonant transitions of C, S, O and Si show four main velocity components, the strong transitions of C, O and Si have two additional subcomponents. These subcomponents might be present also in the weaker transitions but simply not detected in those transitions. \ion{N}{V}$\lambda\lambda$1238, 1242 is detected in a single velocity component at $v_{1*}=$~186\,km\,s~$^{-1}$ (see Fig.~\ref{NV}) which is at slightly higher velocity than comp. 1 of the other absorption lines. The fine-structure lines (Fig.~\ref{finelines}) only show 2 or 3 components which is clearly visible from the only unblended fine structure line, \ion{Si}{II*}$\lambda$1309.

For all host component lines, we assume only turbulent broadening and estimate the values to be $b_\mathrm{turb} = 15 km\,s^{-1}$ in the two narrow components 1 and 4, $25 km\,s^{-1}$ in components 2 and 3 and $20 km\,s^{-1}$ in the additional components 1a and 3a. The values are invoked by the best fits from unblended lines. 
The intervening system at $z=1.478$ is unblended and the multiplet fits for the \ion{Mg}{II}\,$\lambda\lambda$2796, 2803 and \ion{Fe}{II}\,$\lambda\lambda$2586, 2600 doublets give consistent results.

\begin{table*}
\caption{\label{abslines} Velocity decomposition and Voigt-profile fitting of the different components in the host galaxy absorption systems. Metallicities are derived from the total column densities including the ground state and fine-structure transitions of the individual ionic species. As solar metallicities, we used the values from \cite{Asplund04}. The velocities of the Doppler parameters are given in rest frame. Zero velocity is the midpoint of the absorption complex, corresponding to $z=4.048$.}
\centering
\begin{tabular}{l l l l l r l l l l}
\hline\hline
ID & $\lambda_{rest}$ & comp.&  $\lambda_{obs}$ & $v_{abs}$ & log $N$ & $b_{turb}$& log $N_{tot}$ & [X/H]&{\tiny blended}\\
&[\AA]&&[\AA]&[km/s]&[cm$^{-2}$]&[km/s]&[cm$^{-2}$]&&{\tiny with}\\ \hline
\ion{S}{II}& 1250.584	&1	& 6316.51 &169		& 14.51$\pm$0.11	& 15			&15.13$\pm$0.05&--0.86$\pm$0.09&\\
	&				&2 	&6314.41 &69			& 14.52$\pm$0.11	& 25			&	&&\\
	&				&3 	&6312.71 &--11		& 14.71$\pm$0.07	& 25			&	&&\\
	&				&4 	&6308.13	 &--228		& 14.27$\pm$0.19	& 15			&	&&\\[1mm]
\ion{S}{II}&  1253.811	&1 	&6332.81	 &169		& 14.51$\pm$0.11	& 15			&15.13$\pm$0.05&&\\
	&				&2 	&6330.70 & 69			& 14.52$\pm$0.11	& 25 			&	&&\\
	&				&3 	&6329.00 &--11		& 14.71$\pm$0.07	& 25			&	&&\\
	&				&4 	&6324.41 & --228		& 14.27$\pm$0.19	& 15			&	&&\\[1mm]
\ion{S}{II}&  1259.519	&1	&6361.64	 &169	& 14.51$\pm$0.11	& 15	&15.13$\pm$0.05	&& 	{\tiny \ion{Si}{II}$_{1260}$}\\
	&				&2	&6359.52 	 & 69		& 14.52$\pm$0.11	& 25	&				&& 	{\tiny \ion{Si}{II}$_{1260}$}\\
	&				&3	&6357.81	 &--11	& 14.71$\pm$0.07	& 25	&				&& 	{\tiny \ion{Si}{II}$_{1260}$}\\
	&				&4 	&6353.20 &--228		& 14.27$\pm$0.19	& 15			&	&&\\[1mm] \hline
 \ion{Si}{II} &1260.4221 	&1 	&6366.20 &169		& 14.41$\pm$0.09	& 15			&15.23$\pm$0.04&--1.08$\pm$0.07&\\
	&				&1a	& 6365.28& 127		& 12.64$\pm$0.13	& 20			&	&&\\
	&				&2* 	&6363.70 & 51			& 14.66$\pm$0.05	& 25			&	&&\\
	&				&3 	&6362.37 & --11		& 14.79$\pm$0.07	& 25			&	&&\\
	&				&3a	&6361.19	& --67		& 13.71$\pm$0.08	& 20			&	&&\\
	&				&4 	&6357.76 & --228		& 14.52$\pm$0.08	& 15			&	&&\\[1mm]
\ion{Si}{II}& 1304.3702	&1	&6588.18  & 169 	& 14.41$\pm$0.09	& 15		&15.23$\pm$0.04 	&&{\tiny \ion{O}{I*}}\\
	&				&1a	&6387.22& 127	& 12.64$\pm$0.13	& 20		&				&&{\tiny \ion{O}{I*}}\\
	&				&2*	&6585.59	& 51		& 14.66$\pm$0.05	& 25		&				&&{\tiny \ion{O}{I*}}\\
	&				&3 	&6584.21	& --11		& 14.79$\pm$0.07	& 25 			&	&&\\
	&				&3a	&6583.00& --67		& 13.71$\pm$0.08	& 20			&	&&\\
	&				&4 	&6579.44	& --228		& 14.52$\pm$0.08	& 15			&	&&\\[1mm] \hline
\ion{O}{I} & 1302.1685	&1 	&6577.06 &169		& 16.02$\pm$0.53	& 15 			&16.33$\pm$0.29&--1.16$\pm$0.31&\\
	&				&2 	&6574.86	& 69			& 15.72$\pm$0.24 	& 25			&	&&\\
	&				&3 	&6573.10 & --11 	  	& 15.52$\pm$0.47 	& 25			&	&&\\
	&				&3a	& 6571.88& --67		& 14.91$\pm$0.13	& 20 			&	&&\\
	&				&4 	&6568.33& --228		& 15.14$\pm$0.18	& 15 			&	&&\\[1mm] \hline
\ion{C}{II} & 1334.5323	&1 	&6740.52 & 169		& 16.40$\pm$0.39	& 15 			&$>$16.85&$>$ --0.39&\\
	&				&1a	&6739.55& 127		& 13.67$\pm$0.42	& 20			&	&&\\
	&				&2	& 6738.28 & 69			& 15.10$\pm$0.40	& 25 			&	&&\\
	&				&3 	&6736.47 & --11		& $>$16.50		& 25 			&	&&\\
	&				&3a	&6735.22	 & --67		& 14.69$\pm$0.18	& 20			&	&&\\
	&				&4 	&6731.58 & --228		& 16.09$\pm$0.24	& 15 			&	&&\\[1mm] \hline
\ion{N}{V} &1238.821	&1 	&6257.46 &186		& 13.73$\pm$0.06 	& 25 			&13.73$\pm$0.06&---&\\
\ion{N}{V} &1242.804	&1 	&6277.58 &186 		& 13.73$\pm$0.06	& 25 			&13.73$\pm$0.06&&\\ \hline
\end{tabular}
\end{table*}

\begin{table*}
\caption{\label{abslinesfine} Same as Table \ref{abslines} for the fine-structure lines in the host galaxy system that only have 3 components.}
\centering
\begin{tabular}{l l l l l l l l l }
\hline\hline
ID & $\lambda_{rest}$ & comp.& $\lambda_{obs}$ & $v_{abs}$ & log $N$ & $b_{turb}$& log $N_{tot}$&{\tiny blended} \\
&[\AA]&&[\AA]&[km/s]&[cm$^{-2}$]&[km/s]&[cm$^{-2}$]&{\tiny with}\\ \hline
\ion{Si}{II*}	& 1264.7377	&1	&6388.00 	&169	& 14.17$\pm$0.04	& 15 &14.38$\pm$0.03&	\\
			&			&1a	& 6387.08	& 127	& 12.89$\pm$0.14	& 20	&&{\tiny \ion{Si}{II*}$_{1265}$}\\
			&			&2	&6385.87 & 69		& 13.72$\pm$0.07	& 25	&&{\tiny \ion{Si}{II*}$_{1265}$}\\
			&			&3	&6384.16 & --11	& 13.41$\pm$0.05   	& 25	&&{\tiny \ion{Si}{II*}$_{1265}$}\\
			&			&3a	&6382.97 	& --67	& 12.73$\pm$0.08	& 20	&&\\[1mm]
\ion{Si}{II*}	&1265.0020  	&1	&6389.33 	& 169	 & 14.17$\pm$0.04	& 15	&14.38$\pm$0.03&	\\ 
			&			&1a	&6388.41	& 127	& 12.89$\pm$0.14	& 20	&&{\tiny \ion{Si}{II*}$_{1264}$}\\
			&			&2	&6387.21 & 69		& 13.72$\pm$0.07	& 25	&&{\tiny \ion{Si}{II*}$_{1264}$}\\
			&			&3	&6384.16 & --11	& 13.41$\pm$0.05 	& 25	&&{\tiny \ion{Si}{II*}$_{1264}$}\\
			&			&3a	&6384.21	& --67	& 12.73$\pm$0.08	& 20	&&{\tiny \ion{Si}{II*}$_{1264}$}\\[1mm]
\ion{Si}{II*}	&1309.2757	&1	&6612.95 	& 169	& 14.17$\pm$0.04	& 15	&14.38$\pm$0.03&	\\
			&			&1a	&6612.00	& 127	& 12.89$\pm$0.14	& 20	&&\\
     			&			&2	&6610.75 	& 69		& 13.72$\pm$0.07	& 25 	&&\\
			&			&3	&6608.98 	& --11	& 13.41$\pm$0.05   	& 25	&&\\
			&			&3a	&6607.75	& --67	& 12.73$\pm$0.08	& 20	&&\\[1mm] \hline
\ion{O}{I*}		&1304.8576	&1	&6590.64 	& 169	& 14.86$\pm$0.07	& 15 	&14.89$\pm$0.07&\\
			&			&2	&6588.44	& 69		& 13.47$\pm$0.06	& 25	&&{\tiny \ion{Si}{II}$_{1304}$}\\[1mm]
\ion{O}{I**}	&1306.0286   	&1	&6596.55 	& 169	& 14.15$\pm$0.09	& 15	&14.37$\pm$0.07&	\\
			&			&2	&6594.36 	& 69		& 13.98$\pm$0.12	& 25	&&\\[1mm] \hline
\ion{C}{II*}		&1335.6630	&1	&6746.23	& 169	& 15.76$\pm$0.16	& 15 	&15.77$\pm$0.15&	\\
			&			&1a	&6745.26	& 127	& 12.68$\pm$0.71	& 20	&&\\
			&			&2	&6743.98	& 69		& 13.92$\pm$0.06	& 25 	&&\\
			&			&3	&6743.98	& --11	& 14.02$\pm$0.06	& 25 	&&\\
\ion{C}{II*}		&1335.7077	&1	&6746.46	& 169	& 15.76$\pm$0.16	& 15 	&15.77$\pm$0.15&	\\
			&			&1a	&6745.49	& 127	& 12.68$\pm$0.71	& 20	&&\\
			&			&2	&6744.21	& 69		& 13.92$\pm$0.06	& 25 	&&\\
			&			&3	&6742.40	& --11	& 14.02$\pm$0.06	& 25 	&&\\
\hline          
\end{tabular}        
\end{table*}

\begin{table*}
\caption{Absorption line fitting of the intervening system at $z=1.48$}            
\label{abslinesinter}      
\centering                          
\begin{tabular}{l l l l l l l l}        
\hline\hline                 
ID & $\lambda_{rest}$ & comp.& $\lambda_{obs.}$ & $v_{abs}$ & log N & b$_{rest}$ [km/s]		&log N$_{tot}$\\
&[\AA]&&[\AA]&[km/s]&[cm$^{-2}$]&[km/s]&[cm$^{-2}$]\\ \hline    
\ion{Fe}{II} 	& 2586	& 1	& 6412.72	& 25	& 13.27$\pm$0.07&  19.9$\pm$3.4		&13.78$\pm$0.05\\
			& 2586	& 2	&6411.51&--30	& 13.62$\pm$0.07&  14				&\\
\ion{Fe}{II} 	& 2600	& 1	&6446.24& 25	& 13.27$\pm$0.07&  19.9$\pm$3.4		&13.78$\pm$0.05\\
			& 2600	& 2	&6445.03&--30	& 13.62$\pm$0.07&  14				&\\
\ion{Mg}{II} 	& 2796	& 1	&6932.60& 25	& 13.49$\pm$0.08&  19.9$\pm$3.4		&13.95$\pm$0.08\\
			& 2796	& 2	&6931.30&--30	& 13.77$\pm$0.11&  14				&\\
\ion{Mg}{II} 	& 2803	& 1	&6950.40& 25	& 13.49$\pm$0.08&  19.9$\pm$3.4		&13.95$\pm$0.08\\
			& 2803	& 2	&6949.09&--30	& 13.77$\pm$0.11&  14				&\\                
\hline                                 
\end{tabular}
\end{table*}

\section{Velocity components in the host galaxy}
The velocity components in the absorption lines span a range of about 400~km\,s$^{-1}$.  The outer components at 169~km\,s$^{-1}$ (\#1) and $-$228~km\,s$^{-1}$ (\#4) are relatively narrow with $b$= 15~km\,s$^{-1}$ (\#2 \& \#3), whereas several absorption lines of the two components in between are heavily saturated with $b = 25$ km\,s$^{-1}$. In the highest redshift component, which is slightly offset in velocity, the high ionisation \ion{N}{V} line is detected. 

In Paper I we noted the tentative detection of H$_2$ at the redshift of the host galaxy. The position of the H$_2$ transitions are coincident with the redshift of the main absorption component at v $=$ --11~km\,s$^{-1}$ which would then be associated with dense molecular clouds inside the host. All other H$_2$ transitions are be blended with Lyman-$\alpha$ forest absorption lines and can therefore not be identified. The detection of H$_2$ is also consistent with the high metallicity derived for the host of GRB\,060206 \citep{Ledoux03,Petitjean06,Noterdaeme08}. However, the position of the H$_2$ absorption at the velocity of the main absorption component as seen from the metal line transitions might be the reason why we had identified it with H$_2$ transition which would not have been the case had the absorption occured at the velocity of one of the weaker components.

\subsection{Fine structure levels \label{finestruct}}
In Paper I we used the detected fine structure lines to determine the density and temperature for the highest redshift component, assuming that the lines were excited by collisional excitation. However, the VLT/UVES RRM campaign on GRB\,060418 \citep{Vreeswijk07} has shown that at least some the \ion{Fe}{II} and \ion{Ni}{II} fine-structure lines in GRB afterglows are excited by indirect UV pumping from the bright GRB and its afterglow, rather than by collisional excitation. This was also put forward by \cite{Prochaska06} for the spectrum of GRB\,051111.

 \cite{Vreeswijk07} measured strong variation with time of the columns of absorption lines and use this to obtain a direct distance determination of the absorbing material with respect to the GRB, by fitting the columns with indirect UV pumping models. 
Our spectra of GRB\,060206 do not have the required temporal and spectral resolution for such an analysis. We can, however, determine approximate properties for the absorbers assuming that indirect UV pumping or collisional excitation is the way that the fine structure levels are populated. Furthermore, the combined WHT spectrum was taken when a significant optical rebrightening occurred in the afterglow light curve \citep[see the light curve by][]{Curran07} and UV photons from this rebrightening may have played a significant role in the indirect UV excitation of the transitions. In Table \ref{fineprop} we note the ratios between the fine-structure and ground state column densities for the four main components. Due to the uncertainties in the distribution of the column densities between comp. 1 and 1a and 3 and 3a, the densities of the main- and the subcomponent have been added.

\begin{table*}
\caption{Properties of the four velocity systems observed in the WHT spectrum, see 
Section \ref{finestruct}. The exact contribution of the subcomponents 1a and 3a is uncertain. For the ratios, we therefore add the column densities from the sub- to the corresponding main velocity component. For the calculation of $n_{\rm e}$ and $G/G_0$, the ratios derived from \ion{Si}{II*}/\ion{Si}{II} have been used as they have the smallest errors, according to \citet[][Fig. 7 and Fig. 9]{Prochaska06}. \label{fineprop}}
\centering                          
\begin{tabular}{l l l l l l l l}        
\hline\hline                 
Comp.& $v_{abs}$ & \ion{Si}{II*}/\ion{Si}{II}&  \ion{O}{I*}/\ion{O}{I}&\ion{O}{I**}/\ion{O}{I*}&\ion{C}{II*}/\ion{C}{II} & $n_{\rm e}$& $G/G_0$  \\    
& [km s$^{-1}$]& &  & & & [cm$^{-3}$]  &\\
\hline
1+ 1a 	& 148 	&  0.596$\pm$0.029	&  0.07$\pm$0.03& 0.195$\pm$0.004 & 0.23$\pm$0.05	&$400 \pm 50$           & $3\pm0.5 \times 10^5$\\
2 		&  69 	&  0.115$\pm$0.002	&  0.005$\pm$0.001 & 3.24$\pm$0.19 & 0.07$\pm$0.02&$70 \pm 5$ & $6\pm0.15 \times 10^4$\\
3 + 3a	&  $-$39	&  0.047$\pm$0.001 	& ... & ... & $<$ 0.003&$15 \pm 2$          & $2\pm0.15 \times 10^4$\\
4 		&  $-$228 & $<$ 0.005		& ... & ... &... &  $< 2$      &             $< 5 \times10^3$\\
\hline                                 
\end{tabular}
\end{table*}


We can now use these line ratios to estimate the UV radiation field needed to excite the levels, following \cite[][Fig. 7]{Prochaska06}, 
assuming that UV pumping is the dominant excitation mechanism for this component.  The column density ratio [\ion{Si}{II*}/\ion{Si}{II}] indicates a radiation field intensity for the highest redshift component of $G/G_0 = 3 \times 10^5$ (with $G_0 = 1.6 \times 10^{-3}$ erg cm$^{-2}$ s$^{-1}$). Using the ratio [\ion{O}{I*}/\ion{O}{I}] gives very similar values but with a larger uncertainty. A similar strength of the radiation field was found for the component with the second-highest redshift in the spectrum of GRB\,050730 \citep{Delia}, and is somewhat lower than the value found for GRB\,051111 \citep{Prochaska06}.  The required photon field drops to values of $6 \times 10^4$, $2 \times 10^4$ and $\lesssim10^4$ $G/G_0$ for the lower redshift absorbers 2, 3 and 4, the uncertainties for G, however, are quite large (see \cite{Prochaska06}). If all the velocity components in the host are excited through indirect UV pumping by photons of the GRB and afterglow the highest redshift component would be closest to the burst, and the lowest lies furthest away.
Fine-structure lines of \ion{Si}{II*} were also detected in the spectra of LBGs \citep{Pettini02} which might be excited by the UV radiation field by young stars and might therefore be present in any highly star-forming galaxy. The decreasing strength of the lines however are in favour of a well confined UV source like a GRB and its afterglow.

In case we assume collisional excitation being the mechanism responsible for the fine-structure lines, the required densities are 400, 70, 15 and $<$ 2\,cm$^{-3}$ for the systems 1 to 4 (see Table \ref{fineprop}), using the method described in
\cite[][see Fig. 9]{Prochaska06}. These values are comparable to the electron densities found in star forming regions and molecular clouds in our Galaxy and GRB host galaxy star forming regions \citep[e.g.][and references therein]{Thoene06,Wiersema07}. 

We consider indirect UV pumping by the GRB and its afterglow radiation as the most likely scenario for causing the fine-structure lines although we cannot exclude that collisional excitation could play a role. The drop in the inferred radiation field from the higher to the lower redshift components and the lack of fine-structure absorption in component 4 indicates an increasing distance from the GRB from component 1 to 4. In the indirect UV pumping scenario we may interpret the decrease in fine structure line excitation with velocity along the line of sight as a distance effect.

No neutral species with ionization potentials lower than hydrogen are observed in the highest redshift absorber (\ion{O}{I} has an
ionization potential just over 1 Ryd and can therefore be screened by hydrogen), so we cannot set a constraining lower limit to the distance to
the GRB for this system through ionization analysis. If we assume that indirect UV pumping by the afterglow of the GRB is exciting the levels in the highest redshift system, we can derive an order of magnitude estimate of the distance to the GRB \citep[see][]{Prochaska06} using the value of $G$ found above. From the afterglow light curve \citep{Curran07} we calculate the flux received at Earth at the time of the spectrum, using the spectral index derived in Section \ref{extinction}  and the restframe frequency of $1.7 \times 10^{15}$ Hz as in \cite{Prochaska06}. We find a distance of the highest redshift system to the GRB of order one kpc, which is similar to the more  accurate distance found for the absorbing gas through absorption line variability in GRB\,060418 (\citealp{Vreeswijk07}).

\subsection{\ion{N}{V} absorption lines}
\ion{N}{V} has a large ionization potential (77.5 to 97.9 eV) and is therefore assumed to arise only in high temperature or shock environments. It has been frequently detected in GRB sightlines (Prochaska et al., submitted) with large column densities of  log(N) $>$ 14 cm$^{-2}$. They seem to probe cold, narrow gas and often only have small offsets from low ionization and fine-structure lines as we also observe for the spectra presented in this paper. Prochaska et al. argue that the \ion{N}{V} absorption likely arises from photoionization of the immediate ($<$ 10 pc) environment of the GRB through the afterglow, but also shock ionization in starburst winds could in principle play a role and \ion{N}{V} has been observed in LBGs and starbursts \citep{Pettini02, Savaglio02}. Heated gas in the halo traced by other high ionization lines such as \ion{O}{VI}, cannot produce the required high column densities while a connection to shocks from the progenitor wind would imply large offsets which are not observed.

\ion{N}{V} has also been observed in many QSO spectra and in lines of sight towards
stars in the LMC \citep[e.g.][]{caulet96} where these lines are however broader than those observed in QSO and high redshift starbursts. Galactic sightlines towards early type stars \citep[e.g.][]{Welsh05,Sembach01}, in Wolf-Rayet (WR) star winds and galactic outflows. In the solar neighbourhood, it often occurs at the (conductive) interfaces
between evaporating hot and cooler interstellar gas \citep{Slavin89, Welsh05},
e.g. at the boundary of a hot bubble formed by O and B stars with the neutral
ISM, where one would expect to find both high ionization and low ionization
species at comparable velocities. The observed high $b$ parameter is consistent with this picture \citep{Savage} but larger as observed for GRB \ion{N}{V} absorption and the column density is usually log(N) $<$ 14 cm$^{-2}$.

\subsection{The nature of the absorption systems}
Features similar to what we observe in the spectra of GRB\,060206 have been discovered in other GRB spectra such as GRB\,020813 \citep{Fiore05}, GRB\,021004 \citep[e.g.][]{Fiore05}, GRB\,030329 \citep{Thoene06}, GRB\,050730 \citep{Starling05,Chen05, ProchaskaISM}, GRB\,050820 \citep{ProchaskaISM}, GRB\,050922C \citep{Piranomonte07}, GRB\,051111 \citep{Penprase,ProchaskaISM}, GRB\,060218 \citep{Wiersema07} and GRB\,060418 \citep{Vreeswijk07, ProchaskaISM}. The velocity range of these systems span a wide range from 200 to 3000 km s$^{-1}$ and might have different origins according to their widths, ionization states and abundance ratios.

Components with large velocity spreads were sometimes connected to earlier mass losses of the progenitor Wolf-Rayet (WR) star. \cite{vanMarle05} numerically evaluate a scenario in which the progenitor star wind interacts with circumstellar material, including the mass loss of earlier stages in
the progenitor evolution, forming shells of material propagating outwards at a range of velocities. The lower velocity components up to 400~km~s$^{-1}$ match very well with the models by Van Marle et al, but we do not detect the expected components at  2000-3000~km~s$^{-1}$. In the spectra of GRB\,021004, these high velocity components were only detected in \ion{C}{IV} and \ion{Si}{IV} which are outside the usable range of our WHT spectrum. Furthermore, all components of the GRB\,060206 absorption systems show strong low ionization lines which disfavours an origin very close to the burst as the radiation field from the GRB ionizes the surrounding material. \cite{Chen07} conclude from a sample of high-resolution afterglow spectra that high velocity components are absent in 80\% of the spectra and the incidence of \ion{C}{IV} absorbers along DLA sightlines nonnegligible which makes it more likely that those high velocity structures arise in intervening galaxies.

Excluding the connection to the immediate environment of the burst, the absorption features must originate in the host galaxy ISM. Systems with small velocity spreads (v~$<$~250~km~s$^{-1}$) have been found in high resolution spectroscopic studies on QSO absorbers using mainly \ion{Mg}{II} and \ion{C}{IV} systems \citep[e.g.][]{Nestor05, Churchill, Boksenberg03}. These systems are often connected to strong Ly$\alpha$ absorption and consist of a saturated central profile and weaker subsystems at higher velocities. \cite{Churchill} and \cite{Ellison03} suggested a rotating disc together with a halo to explain the different systems. However, as shown in Sec.\ref{extinction}, the extinction in the line-of-sight to the GRB is consistent with zero which makes it unlikely that the GRB was situated behind most of the disc material. Also the hydrogen column densitiy of log$N=20.85 \pm 0.1$ (Paper I) is relatively low for a GRB-DLA \citep{Jakobsson06b} implying a position of the GRB inside or at the near end of the galaxy. Furthermore, fine-structure lines are usually detected only in the component that dominates the optical depth, associated with the gas inside the galaxy and their redshift has therefore been taken as a proxy for the redshift of the galaxy \citep{Prochaska07}.

Another possible explanation for those components is absorption by material from a supergalactic wind \citep{Bond01}. The cumulative supernova explosions within a star forming region can expel gas out into the halo of the galaxy which fragments and produces clumps of material outside the galaxy. These galactic winds have been directly observed in some nearby galaxies (e.g. M82, NGC 3079 and NGC 4945). The inner parts of these superbubbles are relatively hot and show X-ray emission together with optical emission lines, followed by a warm transition phase traced by highly ionized metal absorption lines such as \ion{O}{VI} and \ion{C}{IV}. The outer shells are colder and show low ionized absorption systems \citep[see e.g.][]{Heckman01}. Low ionization absorption features in GRB spectra that can be explained by galactic winds have been found in GRB\,030329~\citep{Thoene06} and GRB\,051111 \citep{Penprase}. \cite{Bond01} and \cite{Ellison03} suggest that the symmetric structures in some \ion{Mg}{II} absorption systems towards QSOs \citep{Bouche05} could also be an indication for a galactic-scale outflow scenario.

If we can transform the radiation field derived from the fine-structure lines into a distance scale for the different components, assuming that G $\sim$ 1/r$^2$ \citep[see][]{Prochaska06} and component 1 to be at $\sim$ 1 kpc (see Sec. \ref{finestruct}), component 2, 3 and 4 would then be at distances of 2.2, 5.5 and $>$ 8 kpc from the burst. Comparing this with the velocity spread of 417 km s$^-1$, the absorbing systems can hardly be caused by the rotation field in an 0.3 L* galaxy with a typical size of a few tens of kpc and a rotational velocity of around 150--200 km s$^{-1}$. However, the distances are derived assuming steady-state conditions for the radiation field, which is clearly not the case for a GRB, and can  only be considered as rough estimates. Therefore, some of the components could indeed be caused by the gas in the main part of the galaxy whereas others are due to an outflow.

Galactic winds are considered to be the main cause for the metal enrichment of the intergalactic medium. Observing this in the spectra of a burst at $z=4.048$, it shows that these mechanisms started quite early in the history of the Universe. To conclusively discriminate between the possibility of a galactic outflow and  absorption within the galaxy, we would need the exact redshift of the galaxy from emission lines as it was possible for GRB\,030329 \citep{Thoene06}. There, the host was a small starburst galaxy with the redshift determined from several emission lines, whereas the Mg I and II absorption systems detected spanning a range of 260 km/s were mostly blueshifted compared to the host emission lines. This indicates some sort of outflow from that galaxy which was interpreted as a sign of a galactic superwind.

\subsection{Comparison with QSO-DLAs}
\begin{figure}
\centering
\includegraphics[width=\columnwidth]{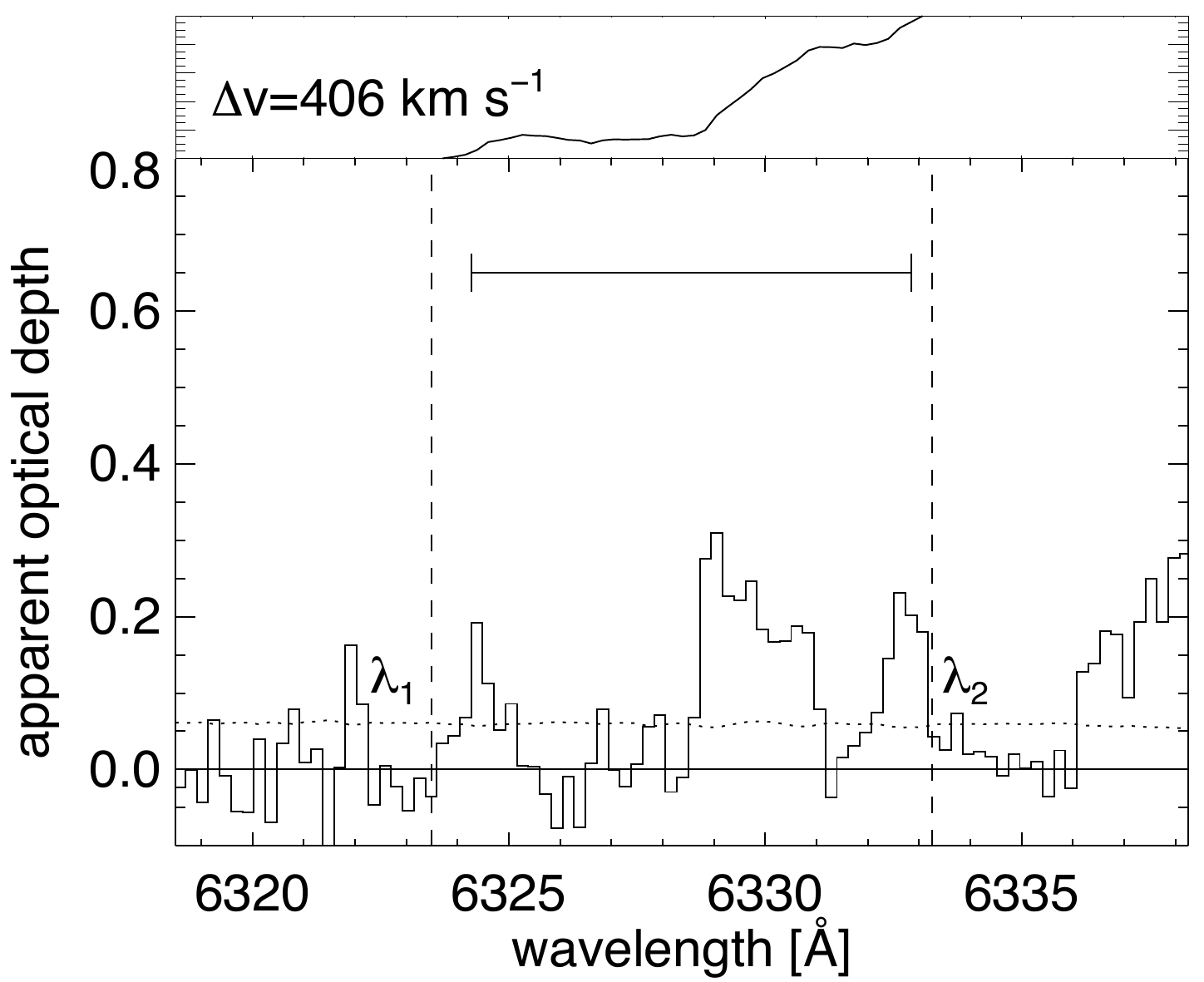}
   \caption{Apparent optical depth distribution of the \ion{S}{II}\,$\lambda$1253 system from which we infer a velocity width of 417 km s$^{-1}$. $\lambda_1$ and $\lambda_2$ indicate 5 and 95\% of the apparent optical depth distribution. The top panel shows the cumulative EW along the line-of-sight}
      \label{width}
\end{figure}

To compare the kinematics with QSO-DLAs we follow the procedure of \cite{Ledoux06} and calculate the line profile velocity width, $\Delta v$, as
$c\left[\lambda(95\%)-\lambda(5\%)\right]/\lambda_0$, where $\lambda(5\%)$ and $\lambda(95\%)$ are the wavelengths corresponding to, respectively, the five and 95 percentiles of the apparent optical depth distribution, and $\lambda_0$ is the first moment (the average) of this distribution \citep[see Fig.~1 of ][]{Ledoux06}. We choose the \ion{S}{II}\,$\lambda$1253 transition as it is a low-ionization transition and the line is not saturated. The apparent optical depth for the line at the derived velocity width is shown in Fig.~\ref{width}. We infer a velocity width of 417 km s$^{-1}$, slightly larger than the distribution of DLAs with comparably high metallicities \citep[Fig.~2 in][see also Prochaska et al. 2007]{Ledoux06}.

This indicates that the kinematics of GRB absorbers is similar to that of QSO-DLAs and hence that the immediate environment of the GRB, e.g. the wind from the progenitor star, is not strongly dominating the kinematics and metallicity of the system. This is consistent with the recent findings of \cite{Vreeswijk07} and \cite{Watson07} that the GRB absorbers probe gas that is more than several kpc away from the GRB site. It also supports our suggestion that the structures observed come from material in the galaxy or from a galactic outflow far away from the GRB.

\section{The host in emission}\label{host}
In the deep $r^{\prime}$ Gemini images of 2700 s, we find a candidate for the host galaxy with 25 $\pm$ 0.2 mag (AB). The host was not detected and the 3$\sigma$ detection limit at the coordinates of the host is {\it H} $>$ 20.6 mag (Vega). The right panel of Fig.~\ref{host} shows the Gemini $r^{\prime}$ band image. In this image, the host appears to have a rather low surface brightness with two slightly brighter knots where the afterglow position falls on top of one of them. 

The HST image resolves the two knots of the possible host from the Gemini images into two objects. One of them is a faint galaxy underlying this position, with the burst offset roughly 0\farcs2 from its centroid (see Fig. \ref{host}). We assume that this is the host galaxy of GRB 060206. This galaxy lies close to two other galaxies, of which the one to the west was blended blended with the actual host galaxy at the resolution of our ground based Gemini image. One of these nearby galaxies may well represent the system seen as the foreground absorbers within our afterglow spectroscopy. The magnitude of the galaxy, measured in an aperture of radius 0\farcs2 is F814W$_{AB}$ = 27.48 $\pm$ 0.19 mag.

   \begin{figure*}
   \centering
   \includegraphics[width=16cm]{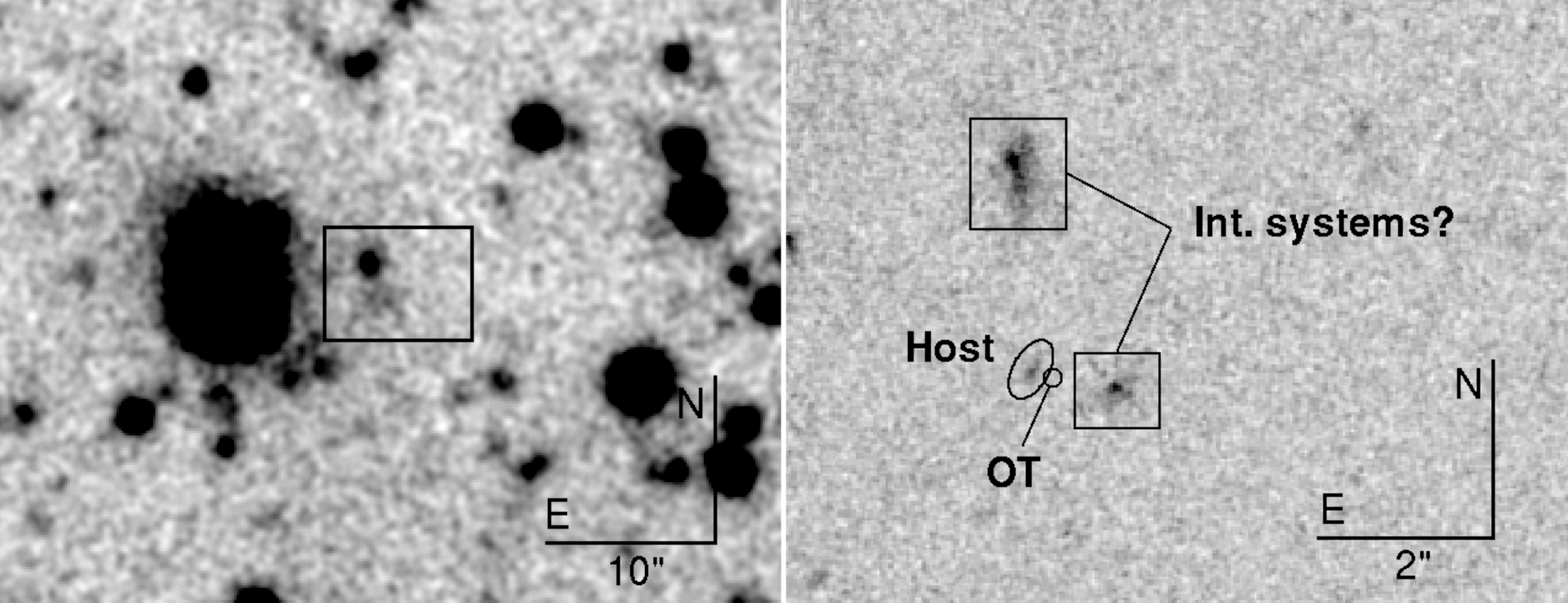}
      \caption{Left panel: Gemini r$^\prime$-band image of the host candidate for GRB\,060206. The box marks the cut displayed in the panel to the right which shows an {\emph HST} images of the host taken with the F814W filter. The OT (circle) lies about 0\farcs2 from a faint object (ellipse) which is most likely the host of GRB\,060206. The two nearby objects (boxes) are likely candidates for the intervening system where the top one corresponds to the object seen inside the box in the Gemini image.}
         \label{host}
   \end{figure*}

The detection of a GRB host galaxy at such high redshift is relatively rare: so far, only three hosts for long GRBs with afterglows at 
$z$$>$3 have been found - GRB 971214 ($z=3.42$) \citep{Kulkarni98}, GRB000131 ($z=4.5$) \citep{Fruchter06} and GRB 030323 ($z$=$3.372$) \citep{Vreeswijk04} despite the fact that 26 GRBs are known with $z$~$>$~3 so far. At a redshift of $z=4.048$, F814W$_{AB}$ = 27.48 corresponds to an absolute magnitude of $-$20.34 mag or roughly 0.3 L*, not taking into account the redshift evolution of the luminosity function. The nondetection of the host galaxy in deep {\it H} band imaging performed with the CAHA 3.5m telescope is consistent with a starforming host with the measured F814W$_{AB}$ magnitude.


\section{An intervening system at $z=1.48$}\label{intervening}

\subsection{Properties of the intervening absorber}
In addition to the systems intrinsic to the GRB redshift, we detect an intervening absorber at  $z=1.48$. It contains two velocity systems of the two doublets \ion{Mg}{II}\,$\lambda\lambda$2796, 2803  and \ion{Fe}{II}\,$\lambda\lambda$2586, 2600 with a separation of $\sim$50 km\,s$^{-1}$. In Paper I, we described a second absorbing system at $z=2.259$. In the reanalysis performed in this paper, we conclude, however, that this system is not real.

In a sample of QSO absorbers, \cite{Churchill} find that the column densities of \ion{Mg}{II} and \ion{Fe}{II} correlate linearly. The column densities we find for both  velocity systems
fall within 1 sigma of the linear correlation of \cite{Churchill}. The $b$
values we find  are very high for both components compared with the sample of
\cite{Churchill}, pointing either to very high temperatures or to more velocity components present that we cannot resolve at our intermediate resolution.
The latter is supported by the correlation that \cite{Churchill} find between the restframe equivalent width of the \ion{Mg}{II}\,$\lambda$2796 line and
the number of velocity components in the absorber, which predicts a higher number of velocity components ($\sim10$) than the two we find in the WHT spectra.

\subsection{Variability of the intervening system?} 
\cite{Hao} claim significant variability in the equivalent widths (EWs) of the \ion{Mg}{II} doublet and the
\ion{Fe}{II}\,$\lambda$2600 line, in the time interval between 4.13 hrs and 7.63 hrs after the burst, covered by 7 low resolution spectra. They fit the lines with Gaussian profiles, and report variation in both the equivalent widths and the relative velocities of the lines. 

The two WHT spectra were taken earlier than those reported in \cite{Hao} with mid-points 1.73 and 2.01 hours after burst. The spectral resolution and signal to noise of the WHT spectra are significantly better than those used by \cite{Hao}. Both the \ion{Fe}{II} and \ion{Mg}{II} lines can be split into two separate velocity components which would be unresolved in  low resolution spectra, but we measure EWs of these two systems combined to compare directly with the \cite{Hao} values. We furthermore analyzed the low-resolution Subaru data, taken around the same time as the spectra reported in \cite{Hao}. 

\begin{figure}
\centering
\includegraphics[width=\columnwidth]{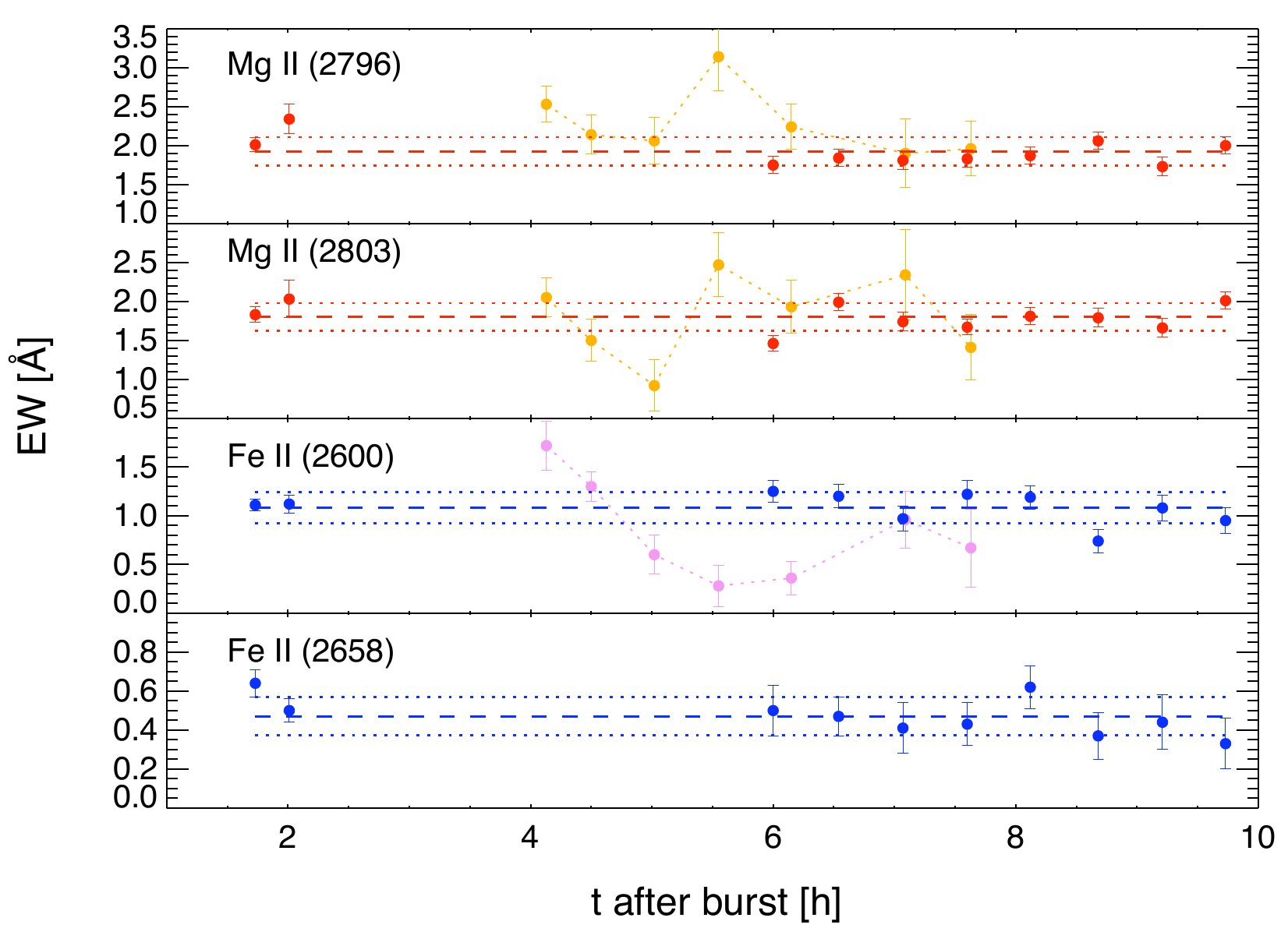}
   \caption{Equivalent widths measured from the WHT and Subaru spectra (red and blue points) and values from \cite{Hao} (orange and violet). In addition, we plot the mean of the WHT and Subaru points (dashed lines) and the 1 $\sigma$ errors (dotted lines) of the mean. Times given are the midtimes of the observations. The plots shows no trend in the evolution of the EW with time as claimed in Hao et al. (2007) and our values are consistent with the mean mostly within 1 $\sigma$.}
      \label{Hao}
\end{figure}

The EW measurements of the \ion{Fe}{II} $\lambda$ 2600 \AA line from both the WHT and the Subaru data show no variations within the errors and are clearly in disagreement with the strong variability in \cite{Hao}. The measurement of the \ion{Mg}{II} lines is complicated by the fact that they are superimposed on the atmospheric B band and the continuum can therefore not be reliably determined. \citep{Hao} removed the absorption band before fitting the \ion{Mg}{II} lines but note that they had a correction of about 25\% of the variation in absorption band caused by the varying airmass during their observations. The apparent variability in the line shape and strength in their \ion{Mg}{II} lines is therefore likely due to inadequate removal of the atmospheric band. We also note that the similar EW of the two components of the \ion{Mg}{II} doublet which should have a ratio of 1:2 indicates that these lines are saturated.
For a comparison between the WHT and the Subaru values and the ones derived by \cite{Hao}, see Fig.~\ref{Hao}.

\begin{table}
\caption{Measurements of the intervening system lines which were reported to vary by \cite{Hao}. The EWs have been measured in the WHT spectra, where we analyzed the two 900s exposures separately and in the Subaru spectra retrieved from the archive. EWs are in the observer frame.}            
\label{whtspecintervening}      
\centering                         
\begin{tabular}{l l l l l l}        
\hline\hline                 
Spec.& t & \ion{Fe}{II}$_{2658}$&\ion{Fe}{II}$_{2600}$& \ion{Mg}{II}$_{2796}$ & \ion{Mg}{II}$_{2803}$\\    
& [h] &EW [\AA]& EW [\AA]& EW [\AA] & EW [\AA]\\
\hline
WHT&1.73& 0.64$\pm$0.07&1.11$\pm$0.06	& 2.01$\pm$0.09& 1.83$\pm$0.10\\
WHT&2.01& 0.50$\pm$0.06&1.12$\pm$0.09	& 2.34$\pm$0.19& 2.03$\pm$0.24\\ \hline                            
Sub.&6.00 & 0.50$\pm$0.13&1.25$\pm$0.11	& 1.75$\pm$0.11& 1.46$\pm$0.10\\
Sub.&6.54 & 0.47$\pm$0.10&1.20$\pm$0.12	& 1.84$\pm$0.11& 1.99$\pm$0.11\\
Sub.&7.07 & 0.41$\pm$0.13&0.97$\pm$0.13	& 1.81$\pm$0.12& 1.74$\pm$0.12\\
Sub.&7.60 & 0.43$\pm$0.11&1.22$\pm$0.14	& 1.83$\pm$0.11& 1.67$\pm$0.10\\
Sub.&8.12 & 0.62$\pm$0.11&1.19$\pm$0.12	& 1.87$\pm$0.11& 1.81$\pm$0.11\\
Sub.&8.68 & 0.37$\pm$0.12&0.74$\pm$0.12	& 2.06$\pm$0.11& 1.79$\pm$0.12\\
Sub.&9.21 & 0.44$\pm$0.14&1.08$\pm$0.13	& 1.73$\pm$0.12& 1.66$\pm$0.12\\
Sub.&9.73 & 0.33$\pm$0.13&0.95$\pm$0.13	& 2.00$\pm$0.11& 2.01$\pm$0.11\\ \hline
\end{tabular}
\end{table}

\subsection{Identifying the absorber in emission}
Detecting intervening galaxies in emission is complicated in the case of QSOs, but the transience of GRBs makes them more suitable to detect intervening systems in emission once the afterglow has faded. While intervening systems can produce very clear absorption systems, the identification of specific absorbers with individual galaxies is difficult \citep[e.g.][]{Vreeswijk2003, Ellison2006}.
The Gemini and {\emph HST} images show several galaxies within a radius of 5 arcsec, with the two most promising candidates $\sim$ 1 and $\sim$2.5 arcsec from the host which corresponds to  distances of $\sim$ 8 and 21 kpc at $z = 1.479$. Photometry of those two galaxies yield magnitudes of F814W$_{AB}$ = 26.1 $\pm$ 0.1 for the one to the right, blended with the host in the Gemini images, and F814W$_{AB}$ = 24.7 $\pm$ 0.1 or $r^\prime_{AB} = 23.9 \pm 0.1$ for the one north of the host. 

At $z = 1.479$, the GMOS $r^\prime$ band and {\emph HST} F814W filter correspond to restframe 2540 \AA\ and 3360 \AA\ and probe restframe UV emission. We estimate the fluxes of the galaxies 1 and 2.5 arcsec north of the host at $\sim$2800 \AA\ to be $\sim$ 0.4 and $\sim$1.2 $\mu$Jy, which corresponds to UV star formation rates of $\sim$ 8 and 25 M$_\odot$yr$^{-1}$, not corrected for extinction, using the conversion from \cite{Kennicutt}. The non-detection of the host or absorber in the {\it H} band is consistent with the intervening absorber candidate being a blue star forming galaxy.

Correlations between parameters of the \ion{Mg}{II} absorption lines and the galaxy that produces them has been studied extensively for QSO absorbers. Recently, \cite{Zibetti} used the large dataset of the SDSS to correlate the QSO \ion{Mg}{II} absorber properties with 
the light of SDSS galaxies at a range of impact parameters (from 10 -- 200 kpc) and found that stronger absorbers can be related to bluer galaxies. The restframe EW of 2.5 \AA{} of \ion{Mg}{II} $\lambda$2796 clearly places it in the category of the strong \ion{Mg}{II} absorbers (though the definition of a strong absorber varies from author to author, but we assume $W_{0}(2796) \simeq 1$\AA), which are generally associated with relatively bright galaxies, with luminosities from $\sim$0.1 to several $L_*$. For two other GRB hosts, GRB\,021004 and GRB\,020813, candidate host galaxies for intervening systems have recently been discovered which seem to lie within a few tens of kpc from the GRB afterglow (Henriksen et al. in prep.). Also GRB\,030429 has a tentative detection of an emission line from the host of the intervening system in the afterglow spectrum \citep{Jakobsson04} and the intervening absorber in the GRB\,020405 spectra could be associated with a neighbouring galaxy with the detection of \ion{O}{II} and \ion{O}{III} at the same redshift \citep{Masetti03}.
 
\cite{Ellison} show that the velocity spread of the \ion{Mg}{II} absorber in combination with the restframe equivalent width 
(the $D$ parameter) supplies an efficient means to select possible DLAs. 
We find $D \sim8$ for our absorber, making it likely that this absorber is a DLA which can be probed by Lyman $\alpha$, which is however still in the UV at z$=$1.48 and could not be observed. The blue, luminous and rapidly star forming galaxy very nearby the GRB host fits in this picture. 

We note that comparison with QSO absorber samples may be further complicated by the recently established over-abundance of \ion{Mg}{II} absorbers on  GRB lines of sight compared to QSO sightlines \citep{Prochter}, for which the cause is yet unclear. This difference is not
present in line of sight detections of \ion{C}{IV}, for reasons not fully understood \citep{Sudilovsky, Tejos}. 

\section{Conclusions \label{conclusions}} 
In this paper we have presented the full set of detected absorption lines in the  medium resolution spectrum from WHT, as well as host galaxy observations with deep
imaging in $r^{\prime}$ and {\it H} band, using Gemini North and the CAHA 3.5m telescope.  The spectra show at least four main velocity components at the GRB redshift, spread
over a range of $\sim400$ km s$^{-1}$. For each component, we detect a series of absorption lines, including several species of fine-structure lines for 3 of the 4 components. Those fine-structure lines are not found in QSO
absorbers, but seem abundant in GRB lines-of-sight \citep[e.g.][]{Vreeswijk04, Starling05b,Prochaska06, fynbob, Kawai06, Penprase}. We measure their column densities through Voigt profile fitting which is complicated by the blending of the lines of different velocity systems, for every velocity component, however, a diagnostic set of absorption lines  can be measured.

The four velocity components show decreasing occupation of their excited fine structure levels as the redshift gets smaller. The column density ratios of [\ion{Si}{II*}/\ion{Si}{II}] indicate that lines in the component with the highest redshift are most likely excited through indirect UV pumping by a strong photon field equivalent to  $\sim 3 \times 10^5~G/G_0$, produced by the GRB and its afterglow. The powerful rebrightening of the afterglow \citep{Wozniak06, Monfardini06, Stanek07} may play an important role in this excitation. If indirect UV pumping through the afterglow is the source of excitation, the highest redshift absorber is in the order of one kpc away from the burst and the other components subsequently at $\sim$2, $\sim$ 5 and $>$ 8kpc. The similarities of the highest redshift component with the values for GRB\,050730 and  GRB\,051111 and the direct detection of absorption line variability of \ion{Fe}{II*}, \ion{Ni}{II*}, but not \ion{Si}{II*}, in GRB\,060418 make an origin of the lines through indirect UV pumping likely. From the fine structure level analysis which provides the relative positions of the velocity components with respect to the GRB, the velocity spread and the low ionization state of the medium, a starburst wind seems an appealing scenario to cause these absorption features. Their nature could be clarified through high resolution (space based) imaging and emission line spectroscopy of the host.


The possible host galaxy is detected at F814W$_{AB}$ = 27.48 $\pm$ 0.19 mag 
and an upper limit of H $\ge$ 20.6 (3 $\sigma$) is achieved which is in agreement with the spectral energy distribution of a blue star-forming galaxy at that redshift. In the image we find several galaxies close to the afterglow position, the closest ones at distances of 1 and 2\farcs5. In the spectrum we detect an intervening absorbers at $z=1.48$ showing a two strong \ion{Mg}{II} and \ion{Fe}{II} absorption components which are likely caused by that bright galaxy. This galaxy is not detected in {\it H} band, consistent with it being a spiral or irregular galaxy, which is also
suggested from the absorption line properties. \cite{Hao} claim significant variability in the equivalent width of the \ion{Mg}{II} doublet and the \ion{Fe}{II}\,$\lambda$2600 line of this absorber. The WHT spectra taken earlier and Subaru spectra taken during the observations of \cite{Hao} do not confirm the variability in equivalent width of these lines or the trend in the variability as a function of time.

The high resolution spectroscopy of GRB\,060206 has been taken with WHT, demonstrating that these detailed line analyses are clearly possible with 4m
class telescopes when a bright afterglow presents itself and intermediate or high resolution spectroscopy is possible.

\begin{acknowledgements}
      We thank the observers and ING staff for performing the reported observations. CT wants to thank Jason Prochaska and Miroslava Dessauges-Zavadsky for the unpublished version of their paper on NV detections in GRB sightlines. 
      
 The Dark Cosmology Centre is funded by the Danish
National Research Foundation. KW, PC and RW thank NWO for support under grant 639.043.302. The authors acknowledge benefits from collaboration within the EU FP5 Research Training Network ``Gamma-Ray Bursts: An Enigma and a Tool" (HPRN-CT-2002-00294). This study is supported by the Spanish Ministry of Science through research projects ESP2005-07714-C03-03 and AYA2004-01515.

Based on observations made with the Nordic Optical Telescope, operated on the island of La Palma jointly by Denmark, Finland, Iceland, Norway, and Sweden and with the William Herschel Telescope, in the Spanish Observatorio del Roque de los Muchachos of the Instituto de Astrof\'{\i}sica de Canarias. Based on observations obtained at the Gemini Observatory, which is operated by the Association of Universities for Research in Astronomy, Inc., under a cooperative agreement with the NSF on behalf of the Gemini partnership: the National Science Foundation (United States), the Particle Physics and Astronomy ResearchCouncil (United Kingdom), the NationalResearchCouncil (Canada), CONICYT (Chile), the Australian Research Council (Australia), CNPq (Brazil), and CONICET (Argentina) and on observations carried out at the Centre Astron\'ominco Hispano Alem\'an (CAHA) at Calar Alto, operated jointly by the Max-Planch Institut f\"ur Astronomie and the Instituto de Astrof\'isica de Andaluc\'ia (IAA-CSIC). Based in part on data collected at Subaru Telescope and obtained from the SMOKA, which is operated by the Astronomy Data Center, National Astronomical Observatory of Japan. 

\end{acknowledgements}

\end{document}